\documentclass[12pt]{article}
\usepackage{amsmath,amssymb,amsfonts,amsthm}
\title{Dynamics of states in the nonlinear interaction regime between a
       three-level atom and generalized coherent states and their non-classical features}

\author{M K Tavassoly$^{1, 2}$,  F. Yadollahi$^{1,3}$
\\
\footnotesize{1- Department of Mathematical Sciences, Yazd University, Yazd, Iran} \\ \footnotesize{2- Research Group of Optics and Photonics,
Yazd University, Yazd, Iran} \\ \footnotesize{3- Department of Physics, Marvdasht branch, Islamic Azad University, Marvdasht, Iran}
\\ \footnotesize{E-mail: mktavassoly@yazduni.ac.ir; ftm.yadollahi@yahoo.com }}

\begin{document}
\maketitle
\begin{abstract}
      The present study investigates the interaction of an equidistant
      three-level atom and a single-mode cavity field that has been initially prepared in a generalized coherent state.
      The atom-field interaction is considered to be, in general, intensity-dependent.
      We suppose that the nonlinearity of the initial generalized coherent state of the field and the intensity-dependent
      coupling between atom and field are  distinctly chosen.
      Interestingly, an exact analytical solution for the time evolution of the state of atom-field system
      can be found in this general regime in terms of the nonlinearity functions.  Finally, the presented formalism
      has been applied to a few known physical systems  such as Gilmore-Perelomov
      and Barut-Girardello  coherent states of $SU(1,1)$ group,
      as well as a few special cases of interest.
      Mean photon number and atomic population inversion will be calculated, in addition to investigating particular
      non-classicality features such as revivals, sub-Poissonian
      statistics and quadratures squeezing  of the obtained states of the entire  system.
      Also, our results will be
      compared with some of the earlier works in this particular subject.
\end{abstract}

{\bf Keywords:}
  Atom-field interaction; Nonlinear Jaynes-Cummings model (JCM); Nonlinear coherent state; Nonclassical state.

{\bf Pacs:} 42.50.Ct, 42.50.Dv, 42.50.Ar, 42.50.-p

\newcommand{\I}{\mathbb{I}}
\newcommand{\norm}[1]{\left\Vert#1\right\Vert}
\newcommand{\abs}[1]{\left\vert#1\right\vert}
\newcommand{\set}[1]{\left\{#1\right\}}
\newcommand{\R}{\mathbb R}
\newcommand{\C}{\mathbb C}
\newcommand{\DD}{\mathbb D}
\newcommand{\eps}{\varepsilon}
\newcommand{\To}{\longrightarrow}
\newcommand{\BX}{\mathbf{B}(X)}
\newcommand{\HH}{\mathfrak{H}}
\newcommand{\D}{\mathcal{D}}
\newcommand{\N}{\mathcal{N}}
\newcommand{\W}{\mathcal{W}}
\newcommand{\RR}{\mathcal{R}}
\newcommand{\HD}{\hat{\mathcal{H}}}


   \section{Introduction}\label{sec-intro}
    Coherent states (CSs) play an important role in quantum optics and modern physics \cite{Ali,kluderbook,perelomovv}.
    Along different kinds of generalization of coherent states \cite{ijmpb},
    nonlinear CSs \cite{Matos Filho} or $f$-CSs $|\alpha,f\rangle$ \cite{Manko} have been introduced and attracted much attention in
    recent decade \cite{Roy}-\cite{Moya-Cessa}
     with the Fock space representation as
    \cite{Manko}:
  \begin{eqnarray}\label{baste nonl}
     |\alpha,f\rangle= \left(\sum _{n=0} ^{\infty } \frac {|\alpha|^{2n}} {n! ( [f(n)]!)^2}\right)^{-1/2}
     \sum_{n=0}^{\infty}  \frac{\alpha^n} {\sqrt{n!}  [f(n)]!}  |n\rangle,
     \qquad\alpha=|\alpha|e^{i\varphi},
  \end{eqnarray}
    where $[f(n)]!\doteq f(1) f(2) ... f(n)$.
    According to this formalism,
    $f$-deformed annihilation and  creation operators,
    respectively are defined as $A=af(n)$ and so $ A^\dag=f^\dag(n)a^\dag$ where $a,a^\dag$ and
    $n=a^\dag a$ are bosonic annihilation, creation and number operator, respectively. Here, the (real) intensity-dependent
    function $f(n)$ is responsible for the nonlinearity of the states.
       On the other side, Jaynes-Cummings model (JCM) is the simplest nontrivial example
       of the atom-field interaction, i.e., a two-level atom and a single-mode radiation
       field \cite{JCM}. As a few recent works in this topics see Refs. \cite{siv, atom-field}.
       The dynamical behaviour
       of the nonlinear atom-field interaction in the presence of classical gravity using the nonlinear coherent states approach discussed in \cite{Naderi}.
       Buzek generalized the JCM \cite{Buzek} and the atom-field
       coupling is considered to be intensity-dependent and supposed that the cavity field be in the Gilmore-Perelomov
       (GP) nonlinear CS of $SU(1,1)$ group \cite{gilmore}. The author showed that, the revivals of the radiation squeezing
       are strictly periodical for any value of initial squeezing parameter \cite{Buzek}. More recently, Koroli {\it et al}
       \cite{Koroli} studied the interaction of an equidistant three-level
       atom (ion), whose dipole moment matrix transition elements between the adjacent atomic levels
       are different, with the GP CS of $SU(1,1)$ group. They showed that, in the three-level
       model with the intensity-dependent coupling, the exact periodicity of the squeezing revivals is violated.

       In the present paper, due to the great interest in the atom-field interactions in the quantum optics,
       we regard the same configuration of three-level atom (ion) has been considered in \cite{Koroli}, however we generalize the
       initial state of the field to the "nonlinear CS" with arbitrary "nonlinearity function"
       $f(n)$. In our formalism, the atom-field coupling is also considered to be intensity-dependent, characterized
       by a function $g(n)$, which  is generally different from $f(n)$. While in Refs. \cite{Buzek, Koroli} the authors have concerned with special nonlinearities,
       our presentation deals, in principle,  with two distinct general nonlinearity functions and interestingly,
       the exact solution is also obtained for the time evolved entire states of the system.

       After finding the explicit solution of the state vector, which evidently depends on $f(n)$ and $g(n)$,
       we will treat the time evolution of the mean photon number and
       atomic population
       inversion, as well as a few non-classicality features such as sub-Poissonian statistics and squeezing of
       the quadratures of the field.
       Therefore, in contrast to the  approach of \cite{Koroli}, that investigated only
       a special system, i.e., Holstein-Primakoff $SU(1, 1)$ coherent states as filed state and $\sqrt n$  as
       the coupling between atom and field, our formalism can be applied to the intensity-dependent
       interaction between the same three-level atom with  arbitrary cavity field
       that is initially prepared in a generalized CS  with known "nonlinearity function $f(n)$". Interestingly, apart from the mentioned
       advantage, as we will observe, our presented formalism allows
       one to work with the general case in which two different nonlinearity functions
       exist, one indicates the initial state of the field ($f(n)$), and the other to the atom-field intensity
       dependent coupling ($g(n)$). Although, it can simply recover the work of Buzek \cite{Buzek}
       and Koroli {\it et al} \cite{Koroli}, as some special cases.
       In the continuation, the
       formalism has been applied to two  classes of generalized CSs, i.e., two distinct representations of $SU(1,1)$
       CSs. In addition, in opposite of the work has been done in
       \cite{Koroli},  in which the authors have taken $f(n)=1/\sqrt{n}$
       and $g(n)=\sqrt{n}$, we choose $f(n)=g(n)=\sqrt{n}$, i.e.,
       the nonlinearity of the initial state is taken to be the same as the intensity-dependent coupling function.
       Finally,  in each of the considered cases, we discuss the numerical results have been shown in
       several figures in detail and  compare with earlier works. Along this procedure,
       some new physical features reveals, which may be emphasized.
       For instance, a typical collapse and revival phenomenon (as a pure quantum mechanical feature)
        in physical quantities such as mean photon number, atomic population inversion
       and Mandel parameter, and the appearance of some
       non-classicality  signs, are specific aspects may be highlighted.

       \section{Hamiltonian of atom-field system}\label{sec-n1}
       We consider the interaction of an equidistant  three-level atom in a cascade
       configuration with
       different dipole moment matrix transition elements between the adjacent atomic energy levels (shown in
       Figure 1) with a quantized single-mode cavity field of frequency $\omega$. The states $|g\rangle,|e_{1}\rangle$
       and $|e_{2}\rangle$ are respectively denoted the ground, first and second excited states. Only the
       atomic transitions between $|g\rangle,|e_{1}\rangle$ and also $|e_{1}\rangle,|e_{2}\rangle$ are dipole
       allowed, but not between $|g\rangle,|e_{2}\rangle$. The Hamiltonian of such an atom-field system is given by:
  \begin{equation}\label{H0,H1}
       H=H_{0}+H_{1},
   \end{equation}
      where $H_{0}$ is the sum of the Hamiltonians of atom and field, i.e., $H_{0}=\hbar \omega_{0}S_{z} + \hbar \omega a^\dag a,$
      $S_{z}=|e_{2}\rangle \langle e_{2} | - |g\rangle \langle g|,$
     denotes the atomic population inversion operator.
     The interaction Hamiltonian between atom and field in (\ref{H0,H1}) in the dipole and rotating wave approximation is given by:
  \begin{eqnarray}\label{H1}
      H_{1} = \hbar \lambda_{1} (a^\dag |e_{1}\rangle \langle e_{2} |+
      a |e_{2}\rangle \langle e_{1} |)
      + \hbar \lambda_{2}(a^\dag |g\rangle \langle e_{1} |
      + a |e_{1}\rangle \langle g |),
  \end{eqnarray}
        where $\lambda_{1},\lambda_{2}$ are the atom-field coupling constants.
        A useful approach to the atom-field interaction problem
        may be found in the "interaction picture" \cite{Zubairy}.
        The Hamiltonian describing the interaction between the mentioned atom-field, in the interaction picture,
        is described by
  \begin{eqnarray}\label{HI}
         H_{I} = \hbar \lambda_{1} (a^\dag |e_{1}\rangle \langle e_{2} | e^{-i\Delta t}+
         a |e_{2}\rangle \langle e_{1} |e^{i\Delta t})
         + \hbar \lambda_{2}(a^\dag |g\rangle \langle e_{1} |e^{-i\Delta t}
         + a |e_{1}\rangle \langle g |e^{i\Delta t})
  \end{eqnarray}
       where, $\Delta=\omega_{0}-\omega$ is the detuning.
       Now, we suppose that the atom-field coupling is intensity
       dependent, expressed by $g(n)$, so the intensity-dependent Hamiltonian in the interaction picture
       is given by:
   \begin{eqnarray}\label{HI}
       \mathbf{\mathcal{H}}_{I} = \hbar \lambda_{1} (R^\dag |e_{1}\rangle \langle e_{2} | e^{-i\Delta t}+
       R |e_{2}\rangle \langle e_{1} |e^{i\Delta t})
       + \hbar \lambda_{2}(R^\dag |g\rangle \langle e_{1} |e^{-i\Delta t}
       + R |e_{1}\rangle \langle g |e^{i\Delta t})
   \end{eqnarray}
     where $R=ag(n)$, $R^\dag=g(n)a^\dag$ and $g(n)$ describes the intensity-dependent coupling
     between atom and field. Notice that, $R$ and $R^\dag$ in (\ref{H1}) have the same structure and
     also meaning of $A$ and $A^\dag$ were described at the beginning of this section, except that a different nonlinearity function is considered.
     Our different notation is only for
     the distinction between the nonlinearity of the initial state of the field  and the intensity
     dependent coupling function. We have assumed that $g(n)$ is a real well-defined
     function with no singularity. The Hamiltonian in (\ref{HI})
     plays a crucial role in determining the subsequent dynamics
     of the quantum states of a variety of (three-level) atom-field systems,
     some of them will be considered in the continuation of the paper.

      \section{ Atom-field state vector}\label{sec-n2}
        Let the atom be initially in the first excited state $|e_{1}\rangle$ and the cavity field
        is prepared in a generalized CS with a nonlinear CS as given in
        (\ref{baste nonl}).
        The state of the atom-field system at $t=0$ can be expressed as:
     \begin{equation}\label{state(t=0)}
       |\Psi (t=0) \rangle = |e_{1}\rangle \otimes |\alpha,f\rangle = \sum _{n=0} ^{\infty }
       C_{n} |e_{1},n\rangle,
     \end{equation}
        where $C_n$ is determined as the expansion coefficients of the states in (\ref{baste nonl}), according to the initial CS has been chosen.
        A useful approach to the atom-field interaction problem
        may be found in the interaction picture \cite{Zubairy}. The Schr\"{o}dinger representation of a state
        vector $|\Psi(t)\rangle$, in terms of its interaction picture representation
        $|\Psi_{I}(t)\rangle$, is given by $|\Psi(t)\rangle = U_{0}(t) |\Psi_{I}(t)\rangle$,
        where $U_{0}(t) =\exp(\frac{-iH_{0}t}{\hbar})$.  Also, $|\Psi_{I}(t)\rangle$ can be obtained from
   $
        i\hbar \frac{\partial}{\partial t} |\Psi_{I}(t)\rangle
        =\mathbf{\mathcal{H}}_{I}|\Psi_{I}(t)\rangle.
    $
        Consequently, by using the above relations, one can straightforwardly find the time evolution
        of the state vector of the coupled atom-field system in the resonance condition
        $\Delta=0$ as follows:
   \begin{eqnarray}\label{state vec sys}
      |\Psi(t)\rangle & =& \sum _{n=0}^{\infty} e^{-i\omega_{0} (S_{z} + {n})t} \; C_{n} \nonumber \\
      &\times &\{\cos (\sqrt{ng^2 (n) + \beta^2 (n+1) g^2 (n+1)}\; \tau) \,|e_{1},n\rangle \nonumber \\
      &-i&\; \frac{\beta \,\sqrt{n+1}\, g (n+1)}{\sqrt{ng^2 (n) + \beta^2 (n+1) g^2 (n+1)}} \nonumber \\
      &\times &  \sin (\sqrt{ng^2 (n) + \beta^2 (n+1) g^2 (n+1)}\; \tau) \,|g,n+1\rangle\} \nonumber \\
      &-i & \sum _{n=1}^{\infty} e^{-i\omega_{0} (S_{z} + {n})t} \; C_{n} \frac{\sqrt{n}\,
      g (n)}{\sqrt{ng^2 (n) + \beta^2 (n+1) g^2 (n+1)}} \nonumber \\ \nonumber \\
      &\times& \sin (\sqrt{ng^2 (n) + \beta^2 (n+1) g^2 (n+1)}\; \tau) \,|e_{2},n-1\rangle,
   \end{eqnarray}
           where $\beta\equiv \lambda_{2}/\lambda_{1}$ and $\tau\equiv\lambda_{1}t$.
           We will call $\tau$ which is the scaled time briefly  as time.
           It is worth noticing that $n$ which is appeared in the exponential functions in (\ref{state vec sys})
           has the operational role (number operator), the same as the role of $S_z$.
           Anyway, this solution is very general and gives the time evolution of a system that
           is involved a single-mode cavity field, has been prepared in a generalized CS,
           coupled with an equidistant three-level atom in a cascade configuration with
           different dipole moment matrix transition elements between the adjacent levels
           $(\lambda_{1}\neq \lambda_{2})$. Two specific cases can be recovered:
           (1) when $\lambda_{2}\rightarrow 0\; (\beta\rightarrow0)$,
           it is equivalent to the single two-level atom; and (2) when
           $\lambda_{2}\rightarrow \lambda_{1} \;(\beta\rightarrow1)$ it corresponds to an equal
           dipole moment matrix transition
           elements between the adjacent levels. This case is indeed equivalent to
           a pair of indistinguishable two-level atoms \cite{Koroli two atom}. The states of the atomic
           pair can be described in the three-level states representation: (i)
           the ground state is equivalent to the case in which both atoms of the
           pair are in the ground state; (ii) the first exited state describes
           the case in which one atom of the pair is in the ground state and another atom is in the exited state,
           and (iii) in the second exited state,
           both atoms of the pair are in the exited state. In addition to the generality of
           our presented formalism, another advantage of
           our formalism is that it contains two distinct nonlinearity functions, i.e.,
           $f(n)$ corresponding to initial state of the field and $g(n)$ which
           determines the intensity-dependent coupling between atom and field.
           So, we can easily recover the results in
           \cite{Koroli}, if we take $f(n)=1/\sqrt{n}$ and $g(n)=\sqrt{n}$ and also in
           \cite{Buzek}, if we take $f(n)=1/\sqrt{n}$, $g(n)=\sqrt{n}$
           with $\beta=0$, as some special cases. Also, we can work with other different possibilities,
           especially the case  $g(n)=f(n)=1$ and $g(n)=f(n)\neq 1$, investigate the output results and compare with \cite{Koroli}.

    \section{Quantum statistics and non-classicality  of
            the introduced state }\label{sec-n4}
       After determining the state vector of the atom-field system at any arbitrary time obtained in
       (\ref{state vec sys}), we are able to investigate mean photon number, atomic population inversion,
        Mandel parameter and squeezing parameters as some non-classicality criteria.

    \subsection{Mean photon number}\label{subsec-n41}

     Using the explicit form of atom-field system $|\Psi(t)\rangle$ in (\ref{state vec sys}),
     the mean photon number can be readily found for any system, with arbitrary
      $f(n)$ and $g(n)$ as follows:
    \begin{eqnarray}\label{mean phot num}
      \langle n\rangle & =&  \sum _{n=0}^{\infty} |C_{n}|^{2}\; n\; \cos^{2}  (\sqrt{ng^2 (n) + \beta^2
      (n+1) g^2 (n+1)}\; \tau) \nonumber \\
      &+& \sum _{n=0}^{\infty} |C_{n}|^{2}\; \sin^{2}  (\sqrt{ng^2 (n) + \beta^2 (n+1) g^2 (n+1)}\; \tau) \nonumber \\
      &\times& \frac{n (n-1) g^2 (n) +\beta ^2 (n+1)^2 g^2 (n+1) }{n g^2 (n) + \beta^2 (n+1)  g^2 (n+1)}.
    \end{eqnarray}

    \subsection{Atomic population inversion}\label{subsec-n42}

      Atomic population inversion, as the expectation value of $S_{z}$, is defined as
  \begin{eqnarray}\label{SZ expectation value}
     \langle S_{z}\rangle=\langle \Psi(t)|S_{z}|\Psi(t)\rangle=|\langle e_{2}|\Psi(t)\rangle|^2-|\langle
     g|\Psi(t)\rangle|^2,
  \end{eqnarray}
     where $|\langle e_{2}|\Psi(t)\rangle|^2$ and $|\langle g|\Psi(t)\rangle|^2$ are the probabilities
     of the presence of the atom in $|e_{2}\rangle$ and $|g\rangle$ states, respectively.
     With the help of  $|\Psi(t)\rangle$ in (\ref{state vec sys}),
     $\langle S_{z}\rangle$ can be written as:
  \begin{eqnarray}\label{SZ expectation value bast}
     \langle S_{z}\rangle  &=& \sum _{n=0}^{\infty} |C_{n}|^{2}  \sin^{2}  (\sqrt{ng^2 (n) + \beta^2 (n+1)
      g^2 (n+1)}\; \tau) \nonumber \\
      &\times& \frac{n g^2 (n) -\beta ^2 (n+1) g^2 (n+1) }{n g^2 (n) + \beta^2 (n+1)  g^2 (n+1)}.
  \end{eqnarray}
     As a special case, in the absence of the radiation field inside the cavity,  namely, the cavity field being in the vacuum state
     $(n=0)$, the quantity $\langle S_{z}\rangle $ is given by:
  \begin{eqnarray}\label{SZ expectation value 0}
     {\langle S_{z}\rangle }_{n=0} = -|C_{0}|^{2} \sin^2(\beta\, g(1)\, \tau).
  \end{eqnarray}
     Recall that, the atom is initially in $|e_{1}\rangle$.
     Therefore, in the absence of a driving field, the atom in the lower state
     $|e_{1}\rangle$,
     cannot excite to the upper state $|e_{2}\rangle$, so ${(|\langle e_{2}|\Psi(t)\rangle|^2)}_{n=0}=0$
     and from (\ref{SZ expectation value}) one has
  \begin{eqnarray}\label{SZ expectation value 01}
     {\langle S_{z}\rangle }_{n=0} = -{(|\langle g|\Psi(t)\rangle|^2)}_{n=0}.
  \end{eqnarray}
    By the comparison of (\ref{SZ expectation value 0}) and (\ref{SZ expectation value 01}), one can arrive at
    $ {(|\langle g|\Psi(t)\rangle|^2)}_{n=0} = |C_{0}|^{2} \sin^2(\beta\, g(1)\, \tau),$
    which  means that in the fully quantum theory of radiation, transition from
    the upper state to the lower state in the
    vacuum of the field becomes possible, known as the spontaneous emission. It is to be noted that this result cannot
    be predicted by semiclassical radiation theory.

     \subsection{ Mandel's $Q$-parameter}\label{subsec-n43}
     To examine the statistics of the states, Mandel's $Q$-parameter is widely used, characterizes the
     quantum statistics of the states inside the cavity.
     This parameter has been defined as  $Q = \frac{\langle n^2 \rangle - \langle n \rangle^2}{\langle n \rangle} - 1$ \cite{Mandel},
     where $\langle n \rangle$ obtained in (\ref{mean phot num}) and $\langle n^2 \rangle$
     may be calculated as follows:
  \begin{eqnarray}\label{mean2 phot num}
      \langle n^2 \rangle & =&  \sum _{n=0}^{\infty} |C_{n}|^{2}\;\left\{ n^2\; \cos^{2}
       (\sqrt{ng^2 (n) + \beta^2 (n+1) g^2 (n+1)}\; \tau)\right. \nonumber \\
       &+&\frac{n (n-1)^2 g^2 (n) +\beta ^2 (n+1)^3 g^2 (n+1) }{n g^2 (n) + \beta^2 (n+1) g^2 (n+1)} \nonumber \\
       &\times&   \left.\sin^{2}  (\sqrt{ng^2 (n) + \beta^2 (n+1) g^2 (n+1)}\; \tau)\right\}.
   \end{eqnarray}
      The Mandel's $Q$-parameter obviously depends on
      the particular choice of $f(n)$ and $g(n)$. It is
      well-known that $Q$
      is positive for classical states (super-Poissonian),
      negative for non-classical states (sub-Poissonian) and vanishes for canonical
             CSs (Poissonian).

    \subsection{Squeezing parameters}\label{subsec-n44}
       Also, we will investigate the squeezing properties of the quadratures of the field.
       For this purpose, we introduce field quadratures as
   $
       X_{1}=\frac{\mathbf{\mathcal{A}}+\mathbf{\mathcal{A}}^\dag}{2},  X_{2}
       =\frac{\mathbf{\mathcal{A}}-\mathbf{\mathcal{A}}^\dag}{2i},
    $
         where $\mathbf{\mathcal{A}}$ and $\mathbf{\mathcal{A}}^\dag$ are the operators
    $
        \mathbf{\mathcal{A}} = a \;e^{i\omega t}, \mathbf{\mathcal{A}}^\dag = a^\dag\;e^{-i\omega t}.
    $
     To study the squeezing properties, we introduce the squeezing parameters:
    \begin{eqnarray}\label{squeezing parameters}
         S_{j} (\tau)= 4 \; \langle(\triangle X_{j})^2\rangle -1,
    \end{eqnarray}
        where $ S_{j} (\tau)$ corresponds to squeezing effect in $X_{j}$ and
        satisfies the inequalities $-1<  S_{j} (\tau) < 0 $. Obviously, to preserve the Heisenberg uncertainty
        relation, when $S_{1} (S_{2})$ is negative, $S_{2} (S_{1})$ should be positive.
        To calculate the parameters $ S_{j} (\tau)$ numerically, one has to  find
        the mean values of the operators: $ a,a^2,a^\dag $ and ${a^\dagger}^2$.
        The following results are easily obtained:
   \begin{eqnarray}\label{mean a,a2}
      \langle a \rangle = e^{-i(\omega t -\varphi)} B_{1} (\tau), \hspace{.2cm}
      \langle a^2 \rangle = e^{-2i(\omega t -\varphi)} B_{2} (\tau),
   \end{eqnarray}
         where
   \begin{eqnarray}\label{B1}
      B_{1} (\tau) &=& e^{-i\varphi}\sum _{n=0}^{\infty}C_{n}^{*} \;C_{n+1}\,\left\{
      \sqrt{n+1}\right.\;\nonumber\\
      &\times & \cos (\sqrt{ng^2 (n) + \beta^2 (n+1) g^2 (n+1)}\; \tau) \nonumber \\
      &\times & \cos (\sqrt{(n+1)g^2 (n+1) + \beta^2 (n+2) g^2 (n+2)}\; \tau) \nonumber \\
      &+&\frac{\sin (\sqrt{ng^2 (n) + \beta^2 (n+1) g^2 (n+1)}\; \tau)}{\sqrt{ng^2 (n)
      + \beta^2 (n+1) g^2 (n+1)} }\nonumber \\
      &\times &\frac{\sin (\sqrt{(n+1)g^2 (n+1) + \beta^2 (n+2) g^2 (n+2)}\; \tau)}{\sqrt
      {(n+1)g^2 (n+1) + \beta^2 (n+2) g^2 (n+2)} }\nonumber \\
      &\times &\left.\sqrt{n+1} g(n+1) [n g(n) + \beta^2 (n+2) g
      (n+2)]\right\}
   \end{eqnarray}
   \begin{eqnarray}\label{B2}
     B_{2} (\tau) &=&e^{-2i\varphi}\sum _{n=0}^{\infty} C_{n}^{*}\; C_{n+2}\,\left\{ \sqrt{n+1}\;\sqrt{n+2} \right.\nonumber\\
     \; &\times & \cos (\sqrt{ng^2 (n) + \beta^2 (n+1) g^2 (n+1)}\; \tau)\nonumber \\
     &\times & \cos (\sqrt{(n+2)g^2 (n+2) + \beta^2 (n+3) g^2 (n+3)}\; \tau) \nonumber \\
     &+&\frac{\sin (\sqrt{ng^2 (n) + \beta^2 (n+1) g^2 (n+1)}\; \tau)}{\sqrt{ng^2 (n) + \beta^2
     (n+1) g^2 (n+1)} }\nonumber \\
     &\times& \frac{\sin (\sqrt{(n+2)g^2 (n+2) + \beta^2 (n+3) g^2 (n+3)}\; \tau)}{\sqrt{(n+2)g^2
     (n+2) + \beta^2 (n+3) g^2 (n+3)} } \nonumber \\
     &\times & \sqrt{n+1} \sqrt{n+2} [n g(n) g(n+2) \nonumber \\ &+&\left.   \beta^2 (n+3) g(n+1)
     g(n+3)]\right\}.
   \end{eqnarray}
      Clearly, $\langle a^\dag\rangle=\langle a\rangle^{\dag}$ and $\langle {a^\dag}^2\rangle=
      {\langle a^2\rangle}^\dag$. It should be noticed that $B_{1}(\tau)$ and $B_{2}(\tau)$ are real values.
      Finally, using the above expressions we arrive at
   \begin{eqnarray}\label{S1}
      S_{1} (\tau)= 2 [B_{0}(\tau)-B_{2}(\tau)] + 4 \;\cos ^2(\varphi) \;[B_{2}(\tau)-B_{1} ^{2}(\tau)],
   \end{eqnarray}
   \begin{eqnarray}\label{S2}
      S_{2} (\tau)= 2 [B_{0}(\tau)-B_{2}(\tau)] + 4 \;\sin ^2(\varphi) \;[B_{2}(\tau)-B_{1} ^{2}(\tau)],
   \end{eqnarray}
     where we have set $B_{0} (\tau) \equiv \langle n \rangle$.

   \section{Some physical realizations of the formalism}\label{sec-n5}
     In this section, we firstly consider the special simple case $f(n)=g(n)=1$ and then nonlinearity functions of CSs of $SU(1,1)$ group are considered.
     \subsection{The special case: $f(n)=g(n)=1$}
     In this subsection it will be constructive to consider  canonical coherent state as the initial state of the field
     and suppose that the coupling between (the three-level) atom and field be independent of intensity, namely $f(n)=g(n)=1$.
     The quantities that have been presented in section 4,
     are plotted versus time for this special case.
     From figures 2-4 it is seen that mean photon number, atomic population inversion and Mandel parameter
     have nearly complete collapses and revivals, at least at intermediate times. This quantities collapses to
     $\langle n \rangle\simeq 63.5$, $\langle S_{z} \rangle\simeq 0.5$ and $Q\simeq 0.01$, respectively.
     Atomic population inversion is positive. This means that the probability of the presence of the atom in $|e_{2}\rangle$ is greater than
     being in $|g\rangle$. Mandel parameter is negative in some time intervals,
     namely the state of the system possesses sub-Poissonian statistics.
     Squeezing parameters versus time is shown in Fig. 5. There is no squeezing in $X_{1}$ and nearly in $X_{2}$, too
     if one ignores the very weak squeezing that may be seen about $\tau=45$.

     \subsection{Nonlinear CSs: CSs of $SU(1,1)$ group}\label{subsec-n51}
     At this stage of the paper, as an example of initial nonlinear CS,
     we will consider the Holestein-Primakoff single-mode realization of $SU(1, 1)$ Lie algebra.
     Before we proceed, it is worth noticing  that, the equivalence of the discrete series representation of $SU(1, 1)$
     state space $\{|\kappa, n \rangle\}_{n=0}^\infty$, with $\kappa= \frac{1}{2},1,\frac{3}{2},2,...\;$,
     and the harmonic oscillator Hilbert space $\{|n \rangle\}_{n=0}^\infty$ is illustrated in \cite{Briff}.
     Based on this recognition, $SU(1, 1)$ CSs have been well established as nonlinear CSs by Ali et al \cite{Ali2}.

     We briefly introduce the GP CSs of $SU(1,1)$ group (sometimes have been
     called Klauder-Perelomov CSs \cite{kluderbook,perelomovv}).
      These states  \cite{Brif} are defined in the interior of the unit disk in the complex plane, centered at the origin.
     The nonlinearity function corresponding to these states is deduced as  $f_{GP} (n,\kappa)
     = 1/\sqrt{n+2\kappa -1}$ \cite{Ali,Roknizadeh-8111,Roknizadeh-042110}.
      In this subsection, we first take $f(n)=g(n)=f_{GP} (n,\kappa)$ in our further numerical calculations.
      Different quantities mentioned in section 4 associated to the atom-field state (\ref{state vec sys}),
      have been plotted versus time
      in figures 6-9 with fixed parameters $\kappa=3/2$, $|\alpha|=0.9$ and $\beta=0.01$.
      Figure 6 deals with the mean photon number.
      A typical fractional collapses and revivals are visible from the
      figure 6a. This figure indicates that the envelope of the oscillations
      fractionally collapses to a fixed value $\simeq 12.3$ and as
      time goes on, the collapsed mean photon number is partly revived.
      The maximum amplitude of the oscillations occurred at $\tau=0$ decreases with time, and the duration
      of the oscillations varies, irregularly.
      Our aim for showing the
      figure in a short time interval is to explain the details of the variation
      of mean photon number in a more apparent fashion, particularly in relation to next figure.
      Atomic population inversion is shown in figure 7. The positivity of this
      quantity at all time
      means that the probability of the presence of the atom in the state $|e_{2}\rangle$ is larger
      than the probability of being in the state $|g\rangle$.
      Similar to mean photon number, fractional collapses and revivals are
      observable from figure 7a.  It is to be noted  that, in this case the atomic
      population inversion collapses to a fixed value $\simeq 0.5$.
      Figures 6b and 7b are the same as figures 6a and 7a, respectively, with the same chosen parameters, except that the
      interval of time are restricted to $20$. A comparison between the figures 6b and 7b shows that they both obey nearly
      the same general pattern and makes one sure that the variations of mean photon
      number and the atomic population inversion are in
      opposite directions, as is expected.
      Figure 8 displays Mandel parameter as a
      function of time. The supper-Poissonian behaviour is observed from the figure, together with a (typical) collapses and revivals.
      This phenomenon occurs for the system under consideration, although not in a clear and regular manner.
      Indeed, Mandel parameter oscillates such drastically with time that the quasi-chaotic behaviour is revealed.
      In figure 9 the curves of squeezing
      parameters $S_{1}$ and $S_{2}$ are shown for $\varphi=\pi/2$.
      Only the negativity of $S_{1}$, which indicates squeezing
      effect in $X_{1}$ quadrature,  is revealed in a finite interval of time.


      At this stage, in contrast to Buzek \cite{Buzek} and Koroli {\it et al} \cite{Koroli} that took $f(n)=f_{GP}
      (n,1/2)=1/\sqrt{n}$ and  $g(n)=\sqrt{n}$,
      we will set $f_{BG}(n)=g(n)=\sqrt{n}$ and follow our numerical calculations.
      The nonlinearity function $\sqrt{n}$ can be associated with
      a particular case of the BG state when one chooses $\kappa=\frac{1}{2}$.
      These states are defined in the whole of the complex plane.
      We emphasize that as in the GP CSs, we consider the Holestein-Primakoff realization of $SU(1, 1)$ Lie
      algebra of BG states.
      Barut-Girardello (BG) CSs of $SU(1,1)$ group \cite{BG}
       are established as the dual pair of GP CSs \cite{Ali2}.
      The nonlinearity function of these states is obtained as
      $f_{BG} (n,\kappa) = \sqrt{n+2\kappa -1}$ \cite{Roknizadeh-8111,Ali2,Roknizadeh-042110}.
     It is easy to check that the operators
     $ K_{-,BG}=af_{BG}(n),  K_{+,BG}=f_{BG}(n)a^\dag, K_{0,BG}=\frac 1 2 [K_{-,BG}, K_{+,BG}] $
     satisfy  the commutation relations
     $[K_{-,BG},$ $ K_{+,BG}]=2K_{0,BG},$ $[ K_{0,BG}, K_{\pm,BG}]=\pm K_{\pm,BG}.$
     where $K_0=n+\kappa$.

      Anyway, choosing $f(n)=g(n)=f_{BG}(n, 1/2)=\sqrt n$, our calculated results have been displayed in figures 10-13.
      The mean photon number and atomic population inversion versus time for the corresponding
      atom-field states have been shown in figures 10 and 11, respectively.
      In the range which the atomic population inversion gets negative values,
      the probability of the presence of the atom in $|g\rangle$ is indeed more than $|e_{2}\rangle$.
      Observing figures 10 and 11, it will be clear that, any increase (decrease) in population inversion
      is simultaneous with a decrease (increase) in mean photon number.
      Also, there are two distinct oscillatory behaviour in these two quantities, one (small) is within the other (large).
      This feature was not appeared in \cite{Koroli},
      when they took $f(n) \neq g(n)$ (figures 1 and 2 of Ref. \cite{Koroli}).
      While there are  some jumps in the mean photon
      number (which simultaneously accompanied by a sudden decrease in the atomic population inversion),
      according to our results small jumps and decrease within great jumps and decrease will be revealed.
      It is worth mentioning that, we continued our numerical results (have not shown here) with
      calculating  the mean photon number and atomic population inversion in a wide interval of time
      and observed that, for both values of the considered $\beta$,
      the quantities are nearly (not exactly) periodic. But, the period of time for the case
      $\beta=0.01$ is larger than the case $\beta=0.1$.
      In figure 12, the Mandel parameter has been displayed as a function of time.
      The results indicate that for $\beta=0.1$ this parameter is almost periodic,
      while its maximum amplitude decreases with time. The sub-Poissonian
      behaviour also occurs in wide intervals of time. For large
      times, full sub-Poissonian statistics will be revealed
      for such system, in both of the presented cases.
      It is worth noticing that, this non-classicality sign has not been observed in \cite{ Koroli}.
      In figure 13, the curves of squeezing parameters
      repeated regularly, but not exactly.
      Only for small intervals of time, squeezing effect is occurred in $X_{1}$ or $X_{2}$ quadrature.

      Also a specific case can be
      regarded, i.e.,  when $f(n)=1$, which means that the initial field is in the standard CS and the
      interaction between atom and field would be intensity-dependent. For instance, we consider $g(n)=f_{GP}
      (n,\kappa) = 1/\sqrt{n+2\kappa -1}$. Our numerical results have been displayed in figures 14-17.
      In figure 14 the mean photon number for the atom-field state
      (\ref{state vec sys}) associated with these particular functions is
      shown.
      Atomic population inversion is  plotted versus time in figure 15.
      A careful observation on the figures 14, 15 leads one to conclude that, when the mean photon number of the
      field is increased (decreased) due to emission
     (absorbtion), the probability of the presence of the atom
      to be in the $|g\rangle$ ($|e_2\rangle$) state increases, consequently the atomic population inversion
      is reduced (increased).
      Figure 16 shows the Mandel parameter, which implies that  it is negative in a wide range of time,
      so that sub-Poissonian behaviour occurs,  repeatedly.
      As shown in figure 17, in small regions
      of time the squeezing effect is observed weakly in $X_{1}$ or $X_{2}$ quadratures, separately.
      Obviously, when $S_{1}(S_{2})$ is negative, $S_{2}(S_{1})$
      have to be positive.

         Summing up the above presented results, two remarkable points may be offered.
         Firstly, the represented numerical
         results plotted in figures 2-5 for the case $f(n)=g(n)=1$
         show that the collapses and revivals occur nearly
         regularly (but not exactly)
         relative to the next ones which contain some kind of nonlinearities. Indeed, in the latter cases the chaotic behaviour will
         be revealed due to the presence of the nonlinearities.
         In general, the Mandel parameter
         and squeezing effect  have not an exact regular
         periodicity, specifically relative to two-level atoms. In addition, as it is expected, the variations of mean photon
         number and the atomic population inversion are in
         opposite directions. As we have explained, this is
         consistence with the physics of the considered interaction.
         Also, the fractional collapses and revivals
         phenomenon, as a well-known non-classicality sign, is seen in the mean photon number,  atomic
         population inversion and Mandel parameter
         in the two groups of figures 6-8.
         The negativity of Mandel parameter in a wide range of time is observed in figures 12 and 16.
         The latter  effect, will become more important, if
         one recalls that the squeezing signature of the
         field is revealed only in a small finite
         interval of time (see figures 9, 13, 17).

           \section{Summary and concluding remarks}\label{sec-n7}
            In summary, we have considered the interaction of an
            equidistant three-level atom in a
            cascade configuration with different dipole moment matrix transition elements
            between the adjacent atomic levels, and a single-mode cavity field that
            is initially prepared in a generalized CS.
            It should be emphasized that, our formalism is presented in a very general regime,
            because it contains two nonlinearity functions: $f(n)$ which characterizes the initial
            state of the field and $g(n)$ which determines the intensity-dependent coupling
            between atom and field. Particularly, $f(n)=g(n)=1$ is equivalent to the case
            in which the initial state of the field is standard CS and the interaction
            between atom and field be independent of intensity.
            Interestingly, we have presented a closed analytical solution for such a non-trivial problem.
            Then, as some physical appearances of the proposed structure, we have investigated the
            mean photon number, atomic population inversion, Mandel parameter and squeezing
            parameters for GP and BG CSs of $SU(1,1)$ group as well as some special cases.
            A few points are remarkable, regarding the presented results.
  \begin{itemize}
     \item
            Unlike the reported work by Buzek in \cite{Buzek} which considered the two-level atom,
            we have not observed  the exact regular
            periodicity of the squeezing parameters, in neither of the chosen nonlinearity functions.
            We also calculated the Mandel parameter and find that
            the exact periodicity which occurs in two-level atom violates.
            These phenomenon are consistence with the reported results in \cite{Koroli}. We investigated and examined this observations for various cases,
            either with the same nonlinearities ($f(n)=g(n)$) or with different nonlinearities  ($f(n) \neq g(n)$).
   \item
         The variation of mean photon
         number and atomic population inversion are in
         opposite directions, which is an expected result,
         in view of the physics of the systems under the considered interaction.
         Also, for the special case $f_{GP}(n)=g(n)=1/\sqrt{n+2\kappa-1}$ in the second group  of figures, the fractional collapses and revivals
         phenomenon in the mean photon number and the atomic
         population inversion, as well as the Mandel parameter
         are new features of our proposal, may be highlighted.
   \item
         Our results confirm that only in the fully quantum theory of radiation, the spontaneous
         emission (transition from the upper state to the lower one) in the vacuum of the field becomes possible.
   \item
         Comparing the figures 2-5, represented the numerical
         results have been plotted for the case $f(n)=g(n)=1$, with next figures one can
         see that the collapses and revivals occur more regularly for
         the former ones relative to others which contain some kind of nonlinearities,
         either in  nonlinearity function of initial state or the intensity-dependent
         coupling. Indeed, in the latter cases the chaotic behaviour will
         be revealed due to the presence of the nonlinearities.
   \item
         Apart from these, the generality and at the same time the simplicity of our proposal allows one
         to apply it to other physical systems with known nonlinearity
         functions, for instance, center of mass motion of trapped ion \cite{Matos Filho},
         photon-added CSs \cite{Sivakumar2}, $q$-CSs \cite{Manko}, deformed photon-added nonlinear coherent states \cite{DPANCS} have been recently introduced by one of us and so on. These are straightforward tasks may be done elsewhere.
         On the other side, Roknizadeh et al introduced a
         Hamiltonian associated with a nonlinear oscillator system,
         based on action identity requirement of nonlinear CSs as follows: $H =  A^\dag A = n f^2(n) \;$
         \cite{Roknizadeh-8111}, upon which one obtains the eigenvalues $ e_{n}  = n f^2(n)$ or equivalently
         $f(n) = \sqrt{\frac{e_{n}}{n}}$. So, obviously the presented approach can be easily applied  to such one-dimensional solvable systems, too.

   \end{itemize}


 \end{document}